\documentclass[11pt]{PoS}
\pdfoutput=1
\usepackage[T1]{fontenc}
\usepackage[utf8]{inputenc}
\usepackage{amsmath,amssymb,amsfonts}
\usepackage{multicol}
\usepackage{wrapfig}
\usepackage{dcolumn,booktabs,colortbl}
\usepackage{defs}

\title{Towards a new determination of the QCD Lambda parameter from running couplings in the three-flavour theory}
\ShortTitle{QCD $\Lambda$-parameter from running couplings in the three-flavour theory}

\author{\hfill\parbox{3cm}{\small\it%
DESY 14-226\\
HU-EP-14/54\\ 
SFB/CPP-14-92\\
}}

\author{%
M.~Dalla~Brida,$^{ab}$ %
\speaker{P.~Fritzsch},$^{c}$ %
T.~Korzec,$^{c}$ %
A.~Ramos,$^{a}$ %
S.~Sint$^{ab}$ and %
R.~Sommer$^{a}$%
\vskip0.25em\\
\llap{$^a$} John von Neumann Institute for Computing (NIC), DESY\\
Platanenallee~6, 15738~Zeuthen, Germany\\
\llap{$^b$}School of Mathematics, Trinity College Dublin, \\
College Green, Dublin 2, Ireland\\
\llap{$^c$}
Institut~f\"ur~Physik, Humboldt-Universit\"at~zu~Berlin, \\
Newtonstr.~15, 12489~Berlin, Germany%
\vskip0.25em\\
E-mail:~\email{mattia@maths.tcd.ie}, \email{fritzsch@physik.hu-berlin.de}, \email{korzec@physik.hu-berlin.de}, 
\email{alberto.ramos@desy.de}, \email{sint@maths.tcd.ie}, \email{rainer.sommer@desy.de} %
}
\abstract{%
\vskip1em
We review our new strategy and current status towards a high precision
computation of the Lambda parameter from three-flavour simulations in QCD. To
reach this goal we combine specific advantages of the Schr\"odinger functional
and gradient flow couplings.
}
\FullConference{The 32nd International Symposium on Lattice Field Theory,\\
		23-28 June, 2014\\
		Columbia University New York, NY}
\begin{document}

\section{Introduction}

In our understanding of the Standard Model the strong coupling $\gbar(\mu)$
plays an essential r\^ole as its parametric uncertainty is one of the dominant
sources of uncertainty in Higgs decays for instance. Using lattice gauge theory
we are in the fortunate position to study QCD for any number of dynamical
flavours $\nf$.  In contrast to any experimental data analysis in high energy
physics, we are able to employ non-perturbative definitions of the strong
coupling by means of any suitable lattice observable $O_{\rm lat}$, such
that~\cite{Flag:2013}
\begin{align}\label{eq:def-gen}
        \gbsq(\mu) &= \lim_{a\to 0} \cc_{O}(a,\mu)\cdot O_{\rm lat}(a,\mu)  \;.
\end{align}
The normalization factor $\cc_{O}(a,\mu)$ guarantees $\gbsq=g_0^2 + \Or(g_0^4)$
at leading order in lattice perturbation theory. Any lattice observable fulfilling
eq.~\eqref{eq:def-gen} defines a renormalization scheme with a different
scale-dependent renormalized coupling. As in QCD conformal symmetry of the
massless Lagrangian is broken on the quantum level, one can define a
renormalization group (RG) invariant, the $\Lambda$-parameter,
\begin{align}\label{eq:def-lambda}
   \Lambda &\equiv  
            \mu\big[{b_0\gbsq(\mu)}\big]^{-{b_1}/({2b_0^2})}\,\ee^{-1/({2b_0\gbsq(\mu)})} 
            \exp\bigg\{\!\!-\!\!\int_{0}^{\gbar^{\vphantom{A}}(\mu)}\!\dd g
            \bigg[\dfrac{1}{\beta(g)}+\dfrac{1}{b_0g^3} -\dfrac{b_1}{b_0^2g}\bigg]\bigg\} \;.
\end{align}
This definition holds for any $\mu$, and its value is trivially
scheme-dependent in the sense that the ratio $\Lambda_1/\Lambda_2$ for two
different schemes can be computed exactly.  It has become standard to quote
$\gbsq$ at the electroweak scale given by the $Z$-boson mass, $\mu=M_{Z}$, in
the intrinsically perturbative $\MSbar$ scheme.

A fully non-perturbative computation of the $\Lambda$-parameter
through lattice computations proceeds in the following way. One chooses an
appropriate definition of a renormalized coupling~\eqref{eq:def-gen} in a
finite-volume renormalization scheme which interlinks the energy scale and the
finite size of the system, $\mu\propto L^{-1}$. Starting at a high energy scale
where perturbation theory is applicable without doubt, say at $\mupt\sim
M_{Z}\sim 100\;\GeV$, one iteratively applies a finite-size rescaling technique
by computing the change of the renormalized coupling from a change of energy
scales (or $L$) by factors of $s=2$ for instance. After $N$ steps one
arrives at some hadronic scale $\muhad=\mupt/s^N\equiv \lmax^{-1}$, where a
connection to large volume (LV) lattice simulations can be established. This
allows to determine $\Lambda$ in terms of some experimentally known hadronic
observable, $\fhad$, used to set the overall energy scale in the LV
simulations, and thus in physical units. For $\Lmsbar$ this strategy decomposes
as follows:%
\footnote{To ease our discussion we work at fixed $\nf$, i.e.,
no quark-tresholds are taken into account.}
\begin{align}\label{eq:Lambda-comp}
        \Lmsbar &= \left[ \fhad \right]_{\rm exp} \times \left[\frac{\Lmsbar}{\Lambda} \right]_{\rm exact}  \times \frac{\Lambda}{\fhad} 
    & &\text{with} &
    \frac{\Lambda}{\fhad} &= \frac{\lmax\Lambda}{\lmax\fhad}
    \;.
\end{align}
The total error on $\Lambda$ is composed of that from the LV scale setting
$(\lmax\fhad)$ and the determination of $\lmax\Lambda$ from the
non-perturbative running in the intermediate, finite-volume scheme. While for
$\nf=2$~\cite{Fritzsch:2012wq} the error of $\lmax\fhad\equiv L_1\fK=0.315(8)(2)$
contributed about $\tfrac{2}{5}$ to the total error of 6\% on ${\Lambda}/{\fhad}$, 
it will become negligible in the near future due to new
developments~\cite{MattiaDB:Lat14}. The current world average(s) for
$\alpha^{(5)}_{\rm s}(M_{Z})$ are $0.1183(12)$ from PDG~\cite{Beringer:1900zz}
using experiments only, and $0.1184(12)$ from the average of present lattice
determinations~\cite{Flag:2013}. This 1\% error translates into an error of
about 6\% in $\Lmsbar^{(5)}$. 
For our new estimate of $\Lambda^{(\nf=3)}/\fhad$, to be derived from LV lattices
within the current CLS effort~\cite{Bruno2014}, it is thus worthwhile to aim for
an accuracy of about 4\% or better.

\section{Why choose a new strategy?}

We have seen that by mainly controlling the accuracy of the non-perturbative RG
running, we can improve the determination of $\Lambda$. In the past, the
Schr\"odinger functional (SF) coupling $\gbsf$~\cite{Luscher:1992an} has been
the most useful definition of a finite-volume renormalization scheme compatible
with the strategy behind eq.~\eqref{eq:Lambda-comp}. 
A more recent development that can be used along the same lines, is given by
one of the many running coupling definitions employing the Yang--Mills gradient
flow (GF)~\cite{Narayanan:2006rf}, $\gbgf$ in short. Initially introduced
in~\cite{Luscher:2010iy} it has been studied in a finite-volume setup
in~\cite{Fodor:2012td,Fritzsch:2013je,Ramos2014}.  Especially, the apparently
much better noise-to-signal ratio of $\gbgf$ raises hope to significantly
increase the accuracy in the RG running. To what extent this statement complies
with the renormalization group running covering two orders of magnitude in
energy scales, we will see below. First we start with a general comparison of
SF and GF couplings, both defined with Dirichlet boundary conditions in time.
A short summary is given in table~\ref{tab:compare}.
\begin{table}[t]
  \small
  \centering
  \begin{tabular}[]{llll}\toprule
    topic                                     & SF coupling                   & GF coupling                     & remark                      \\\midrule
    \sc Definition                            & $\gbsq_{\rm SF}(L) = k \langle\tfrac{\partial\Gamma}{\partial\eta}\rangle^{-1}_{\eta=0}$ 
                                              & $\gbsq_{\rm GF}(L) =   \langle t^2E\rangle /\mathcal{N}$        &                                 \\
    \sc SF boundary field                     & $\ne 0$                       & $=0$                            &                                 \\[0.4em]  
    \sc PT matching $\sim$ 64\,GeV            & 2-loop                        & tree-level                      & GF @ higher energies            \\[0.4em]
    \sc typical \# meas.                      & O(100\,000)                   & O(1000)                         & $\Delta\gbsq_{\tiny\rm SF}\!\simeq\!\Delta\gbsq_{\rm GF}$ \\
    \hskip2em%
    ($\tau_{\rm int}$, $\mathcal{V}$, \ldots) &                               &                                 & $L\approx\!0.4\,\fm$ \\[0.4em]
    \sc cutoff effects                        & mild                          & rather large (so far)           &                                 \\
                                              & 2-loop improvement            & tree-level improvement          &                                 \\
                                              & $\Rightarrow {L}/{a}=6\ldots12$   & $\Rightarrow {L}/{a}=8\ldots16$      & controlled $a\to 0$             \\[0.4em]
    \sc $\Delta \gbsq/\gbsq$                  & $\sim \gbsq$                  & const.                          & for fixed \#meas.               \\
    \sc $\Rightarrow \Delta L/L$              & const.                        & $\sim \text{const}/\gbsq$       &                                 \\
    \bottomrule
  \end{tabular}
  \caption{General comparison of gradient flow and SF running coupling schemes.}
  \label{tab:compare}
\end{table}

The SF coupling, $\gbsf(L)$, is defined as the response to a variation of the
(QCD) action about a non-vanishing Yang--Mills background field imposed through
boundary conditions at Euclidean time $x_0=0,T$ $(T=L)$~\cite{Luscher:1993gh}.
As such it is entirely sensitive to the physical extent $L$. The coupling
$\gbgf(L)$ on the other hand is also sensitive to short distances since it is
given by the Yang--Mills energy density defined with vanishing background field
at finite gradient flow time $t=(cL)^2/8$ for some fixed constant $c\in
[0.25,0.5]$~\cite{Fritzsch:2013je}. While for both definitions the tree-level
normalization $\cc_{O}(a,\mu)$ is known for different lattice
actions~\cite{Luscher:1993gh,Fritzsch:2013je,Ramos2014}, cutoff effects are
mild for $\gbsf$ but large for $\gbgf$.%
\footnote{A consistent Symanzik improvement to reduce cutoff effects in
gradient flow observables and thus $\gbgf$ has been proposed
in~\cite{RamosSint:Lat14}.}
Additionally, for the SF coupling perturbative improvement is known up to
2-loop order but unknown for the GF coupling. Both facts together with
additional details such as the respective integrated autocorrelation time,
statistical variance and so on, influence the precision of the continuum limit
of the lattice step-scaling function $\Sigma$,
\begin{align}\label{eq:ssf}
        \sigma(u) &= \lim_{a\to 0} \Sigma(u,a/L) \;, 
        &
        \Sigma(u,a/L) &= \left.\gbar^2(2L)\right|_{\gbar^2(L)=u,Lm=0} \;.
\end{align}
It has to be computed numerically at chosen values of
$u_{n}\equiv\gbar^2(L_{n})$, $n\in\{1,\ldots, N\}$ in order to cover the energy
range under consideration. From past experience we expect that lattice sizes
$L/a<12$ are sufficient to control the continuum extrapolation of
$\Sigsf(u,a/L)$. On the other hand lattices with $L/a>8$ are needed to achieve
an equivalent accuracy in the computation of $\Siggf(u,a/L)$.  At a fixed
volume with $L\approx 0.4\,\fm$, and choosing $c\equiv \sqrt{8t}/L = 0.3$ such
that $\gbsf^2(L)\simeq\gbgf^2(L)$, a numerical study has shown that in order to
achieve the same accuracy in both couplings one needs 100 times more
measurements of $\gbsf^2$.
\begin{figure}[t]
        \centering
        \includegraphics[width=0.50\textwidth]{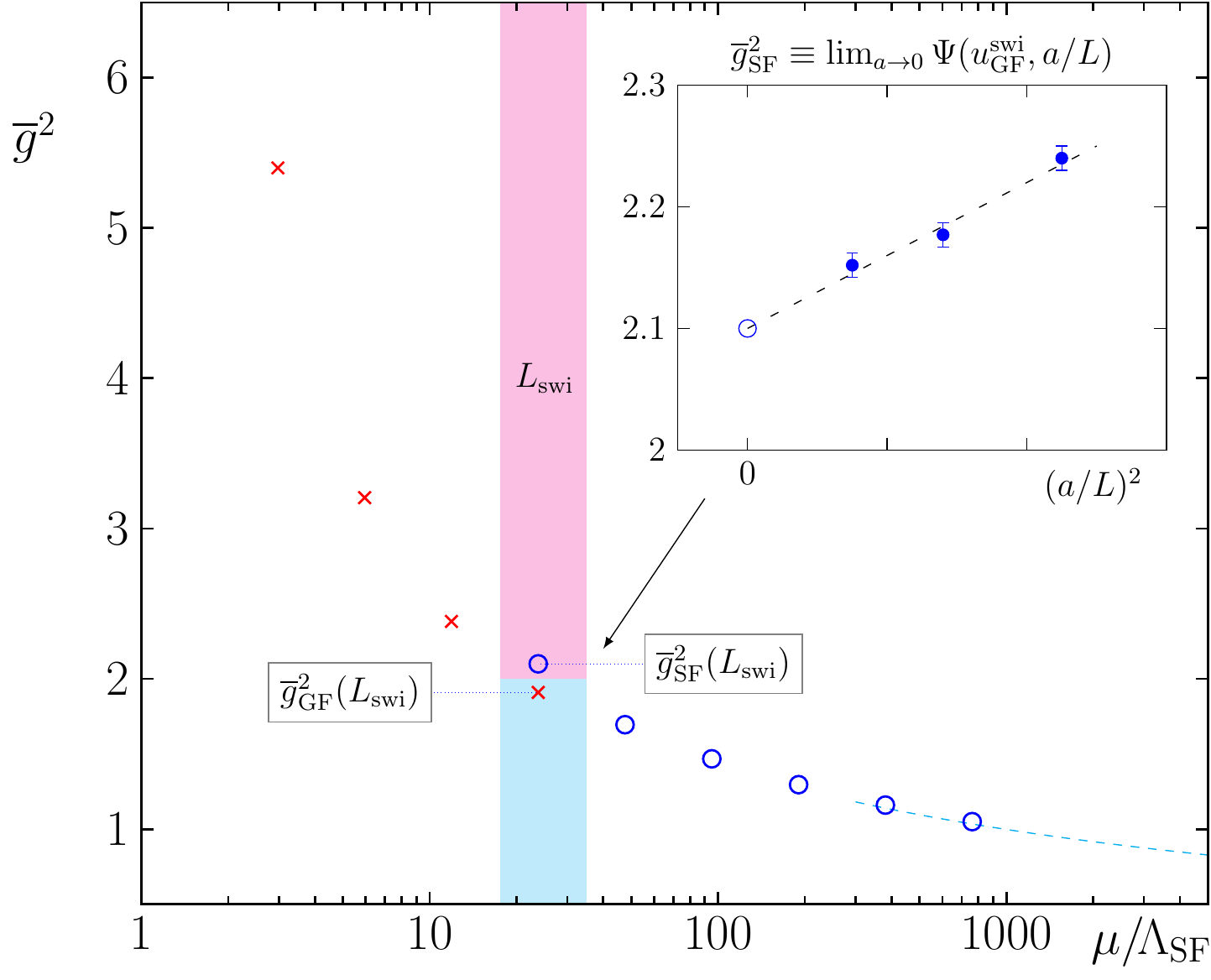}
        \hspace*{1em}
        \raisebox{3.5pt}{\includegraphics[width=0.45\textwidth]{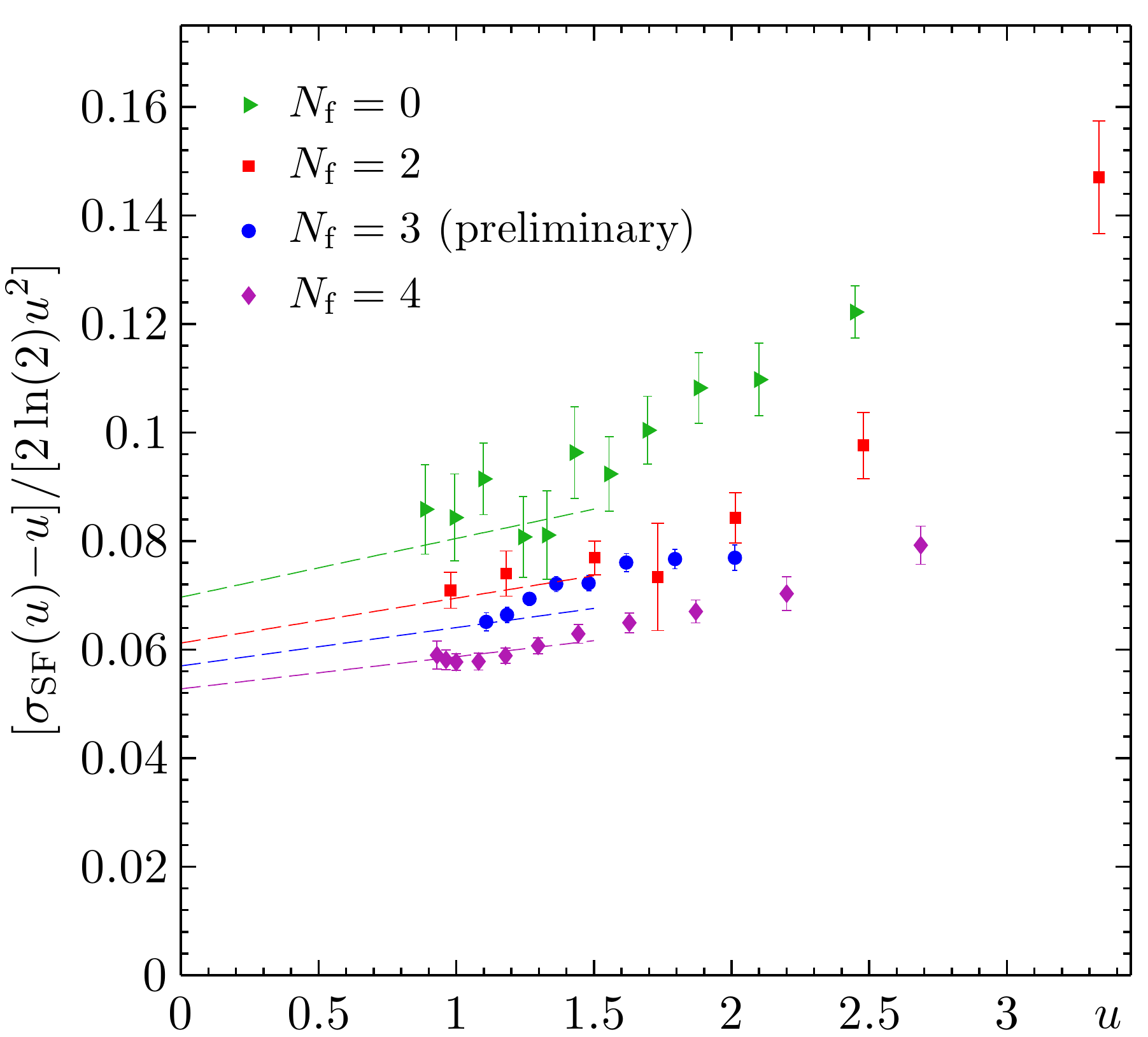}}
        \vskip-0.5em
        \caption{({\em Left}) Sketch of our proposed strategy to non-perturbatively 
                              switch running coupling schemes at an intermediate 
                              energy scale $\mu_{\rm swi}=1/L_{\rm swi}$.
                 ({\em Right}) Data of SF step-scaling functions for different 
                              dynamical flavour content, 
                              $\nf=0,2,3,4$~\cite{Luscher:1993gh,DellaMorte:2004bc,Tekin:2010mm},
                              and its leading asymptotic form $b_0+\Or(u)$ (dashed lines).
                }
        \label{fig:strategy}
\end{figure}
Finally, the two couplings show a very different leading scaling behaviour of
the relative error $R_{\gbsq}\equiv\Delta\gbsq/\gbsq$ which directly translates
via the RG into $R_{L}\equiv\Delta L/L = -\Delta\gbsq\big/2\gbar\beta(\gbar)$.
Considering only the dominant contribution, $\beta(\gbar)\simeq -b_0 \gbar^3 $,
this results into an error of $R_{L}= R_{\gbsq}\big/ 2b_0\gbsq$, with
$(2b_0)^{-1}\approx 9$ in three-flavour QCD. With a behaviour of $R_{\gbsq_{\rm
SF}}\sim\gbar^2$ and $R_{\gbsq_{\rm GF}}\sim\text{const}$, a relative scale
uncertainty of $\Delta L/L\sim\text{const}$ follows for the SF coupling and
$\Delta L/L\sim 1/\gbar^2$ for GF couplings at a fixed number of independent
measurements.

Every step in the finite-size scaling method contributes to the total
error budget of the RG running, giving as first rough estimate
\begin{align}\label{eq:budget}
        \left( \frac{\Delta[\lmax\Lambda]}{\lmax\Lambda} \right)^2
        &\simeq \sum_{n=1}^N \left( \frac{\Delta L}{L}\Big|_{u_{n}} \right)^2 \;,
        &
        u_{n} &= \sigma^{-1}(u_{n-1}) \;,
        &
        u_{0} &= \gbsq(\lmax) \;.
\end{align}
From this point of view it should be clear that at constant effort, the SF
coupling accumulates approximately the same error in each step while the GF
coupling definition shows an error that is steadily growing towards the high
energy regime. Of course, if computational costs are irrelevant this problem
could be solved by brute force calculations. So far we have discussed both
schemes under the aspect of statistical accuracy only, but there are more
points to be taken into account, for instance the matching to PT at high
energies. For $\gbsf$ the matching coefficients in 
$\gbsq_{s}=\gbsq_{s^\prime}+\sum^{\ell}_{k=1}\chi^{(k)}\gbar^{2k+2}_{s^\prime}$
are known up to $\ell=2$ loop order~\cite{Bode:1999sm} while no coefficient is
known for $\gbgf$ so far. Even with a given 1-loop coefficient, $\gbgf$ still
has to be matched at higher energies to control this step equally well,
and thus increases the costs and statistical error in the RG running.
Additional points will be elucidated in more detail in a forthcoming
publication.

Having the global aspect of our strategy in mind we have to conclude that the
GF coupling scheme is advantageous at hadronic scales while the SF coupling
scheme still is to be favoured at high energy scales.

\section{Changing the standard strategy}

From the facts presented in the previous section the optimised way to improve
the computation of $\Lambda$ seems to be to combine both schemes. As depicted
in the left panel of Fig.~\ref{fig:strategy}, one computes the running of
the chosen gradient flow coupling at low energies up to an energy scale of
$\muswi=1/\lswi\equiv s^K/\lmax$:
\begin{align}\notag
        \ugf_{k} &= \gbgf^2(\lmax/s^k)  \;, \quad k\in\{0,1,\ldots,K\}\;,
        &
        \ugf_{0} &= \gbgf^2(\lmax) \;,  \\\label{eq:GFsteps}
        \siggf\left( \gbgf^2(L/s) \right) &= \gbgf^2(L) \;,
        &
        \ugf_{\rm swi} &= \gbgf^2(\lmax/s^K)\;.
\end{align}
At the scheme-switching scale $\lswi$ one has to set up additional simulations 
defined through a line of constant physics, $(\gbgf^2(\lswi),\lswi
m)=(\ugf_{\rm swi},0)$, in order to compute the SF coupling via%
\begin{align}\label{eq:LCPswi}
        \usf_{\rm swi} &= \lim_{a\to 0} \Psi(u,a/L) \;,
        &
        \Psi(u,a/L) &= \left.\gbsf^2(L) \right|_{u=\ugf_{\rm swi},\lswi m=0 } \;,
\end{align}
as accurately as possible. From here on one continues in the new scheme
\begin{align}\notag
        \usf_{m} &= \gbsf^2(\lswi/s^m)  \;, \quad m\in\{0,1,\ldots,M\}\;,
        &
        \usf_{0} &= \gbsf^2(\lswi) \;,  \\\label{eq:SFsteps}
        \sigsf\left( \gbsf^2(L/s) \right) &= \gbsf^2(L) \;,
        &
        \usf_{\rm PT} &= \gbsf^2(\lswi/s^M)\;,
\end{align}
with $M=N-K$ if the same scale difference is to be covered as originally
anticipated by $\mupt^{-1}=\lmax/s^N$, and where ultimately the connection to
the perturbative running can be safely established.
We remark that we could have started our discussion also at high energies,
i.e., reversing the strategy and interchanging SF and GF in
eq.~\eqref{eq:LCPswi}. In fact, carrying out both procedures might help to 
increase control over this non-perturbative scheme-switching step.

\section{Present status}

\begin{figure}[t]
        \centering
        \includegraphics[width=0.45\textwidth]{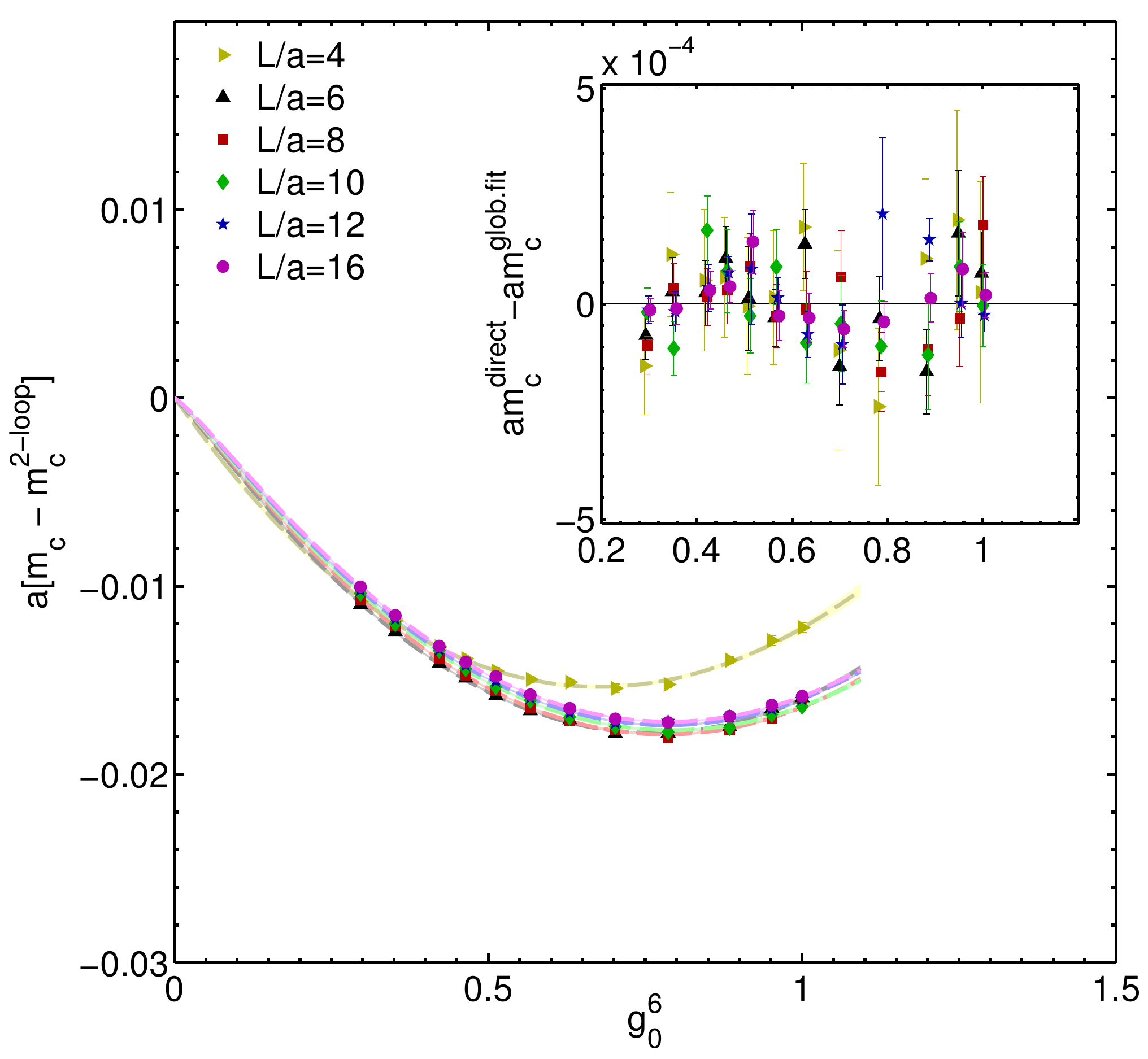}\hfill
        {\includegraphics[width=0.498\textwidth]{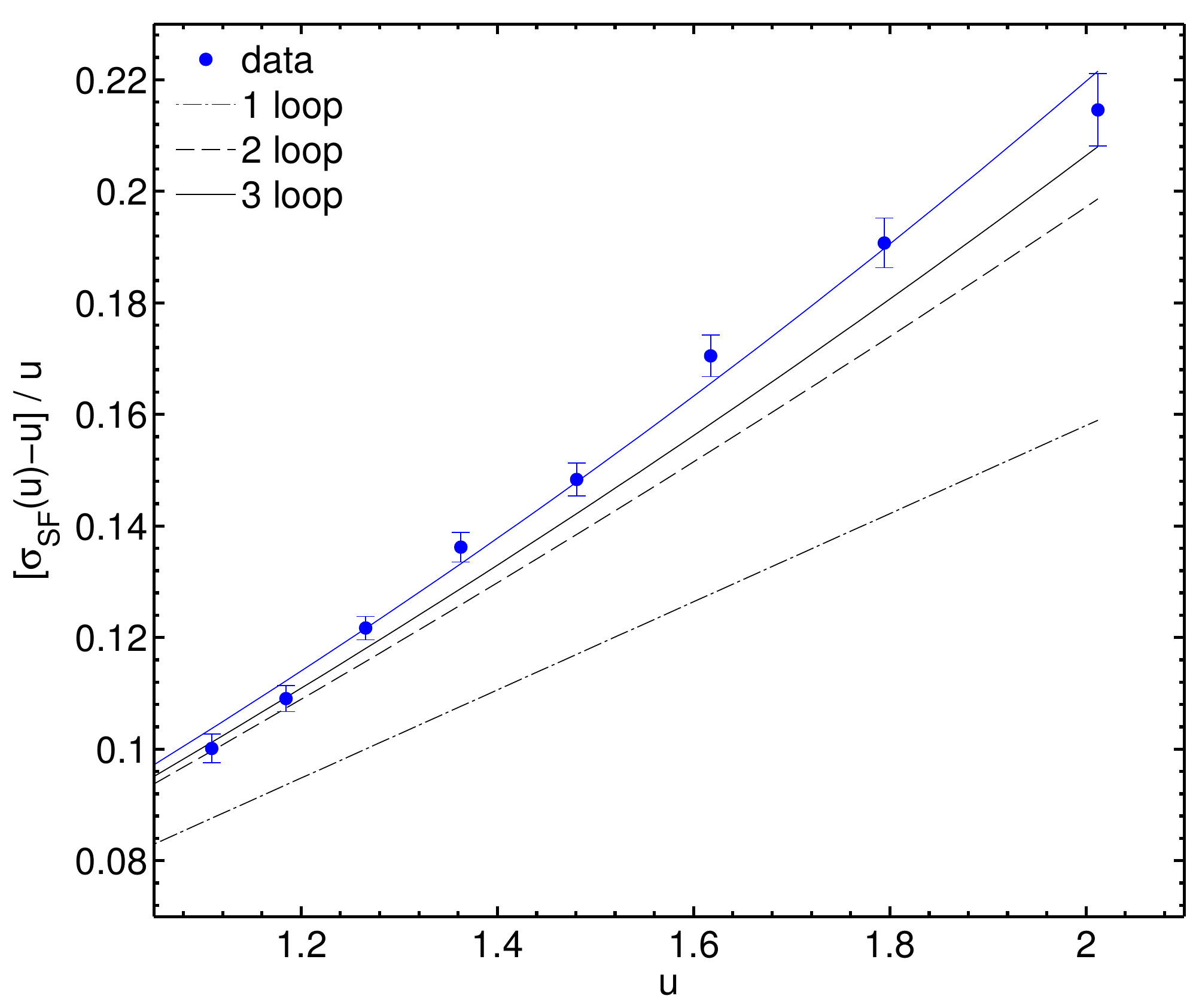}}
        \vskip-.75em
        \caption{({\em Left}) Data and interpolating functions for critical 
                  line $m_{\rm crit}(g_0^2,a/L)$ along constant $L/a$. 
                  ({\em Right}) Non-perturbative continuum step-scaling function 
                  $\sigma_{\rm SF}(u)$ vs. perturbative evaluation.
                }
        \label{fig:run_gSF}
\end{figure}

For various practical reasons we have started with the running of the SF
coupling from high to intermediate energies, reaching couplings as large as
$\usf=2$ within our new strategy. We are working in a massless renormalization
scheme with $\nf=3$ quark flavours and thus need to tune to vanishing PCAC
quark mass, $Lm=0$, for plaquette gauge action and non-perturbatively $\Or(a)$
improved Wilson quarks. This has been done for lattice sizes $L/a=4,\dots,16$
and bare gauge couplings $6 \le \beta\equiv 6/g_0^2  \le 9$. To minimise tuning
efforts in subsequent projects, we have done this more precisely than in the
past, allowing us to smoothly parameterise the critical line
$\mcrit(g_0^2,a/L)$ on the considered lattice sizes for $0\le g_0^2 \le 1$.
Employing the perturbative 2-loop formula for $\mcrit(g_0^2,a/L)$, we use the
fit ansatz
\begin{align}\label{eq:m_crit}
        a\mcrit(g_0^2,a/L) &= [a\mcrit(g_0^2,a/L)]_\text{2-lp} + k_1(a/L)\cdot g_0^6
                            + k_2(a/L)\cdot g_0^8 +   k_3(a/L)\cdot g_0^{10}
\end{align}
to describe our data, c.f. left panel of Fig.~\ref{fig:run_gSF}. The difference
between a direct fit for each $L/a$ and global fits with varying coefficient
functions $k_1,k_2,k_3$ serves as a quality criteria for our interpolating
function. We know that for the accuracy we are aiming at, any additional 
uncertainty from a tuning to a line of constant physics becomes 
\emph{negligible}, if $|Lm|<0.005$ is satisfied. Our final estimates fulfill
$|Lm|<0.001$.

\begin{wrapfigure}{r}{0.5\textwidth}
  \vspace{-4em}
  \begin{center}
        \includegraphics[height=8cm]{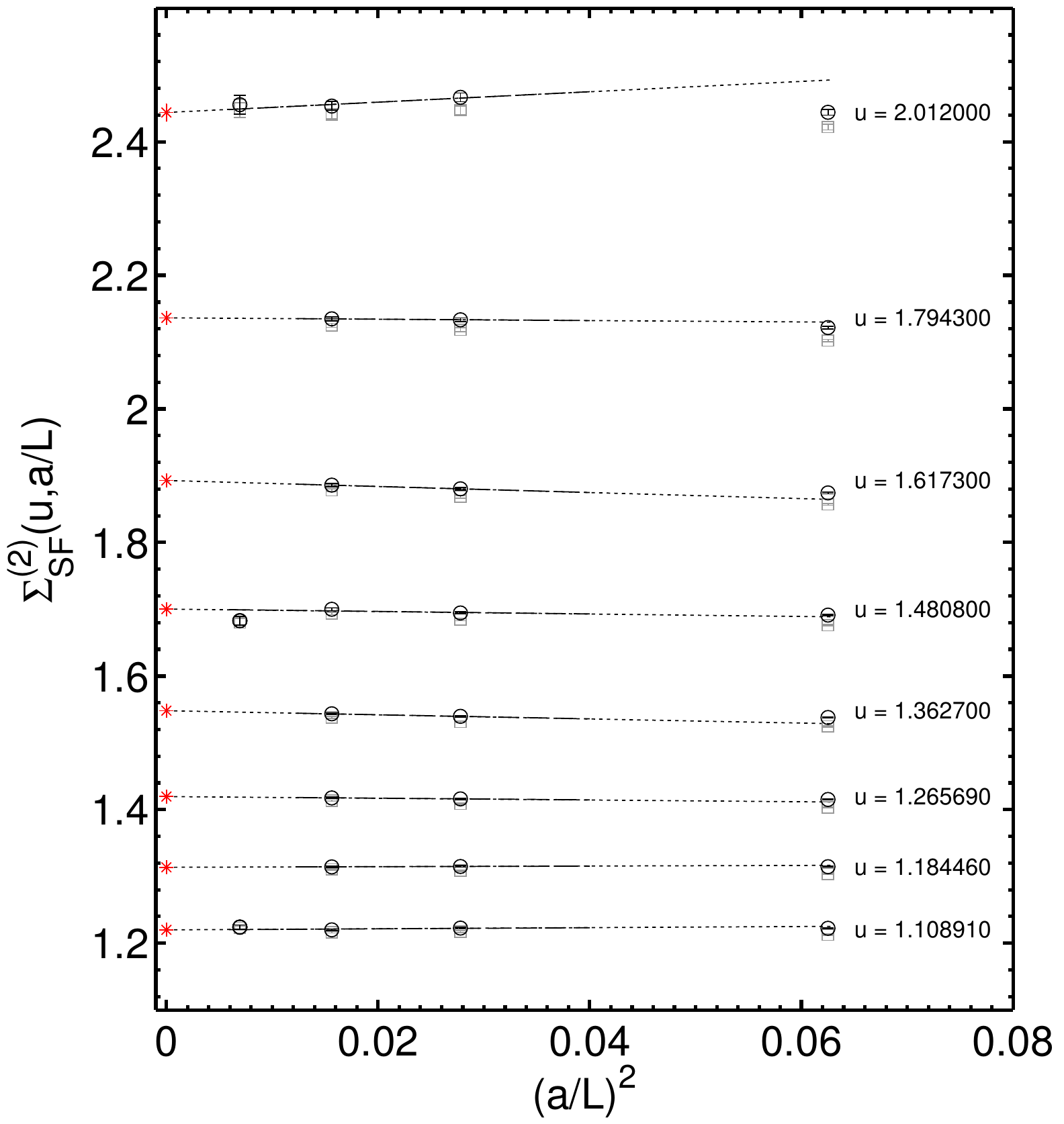}
  \end{center}
  \vspace{-2em}
  \caption{Local (preliminary) continuum extrapolations of
           $\Sigsf^{(2)}(\usf,a/L)$ using $L/a\ge 6$ lattices.
          }
\label{fig:CL}
\end{wrapfigure}

To compute the SF step-scaling function $\sigsf(u)$ we aim for simulations with
$1.1\le \usf \le 2$, uniformly distributed in $1/\usf$. For $L/a\in\{4,6,8\}$
we use 8 different $\beta$ values but simulate only three for $L/a\in\{10,12\}$
in order to minimise computational costs in the finite-size rescaling step
$L\to 2L$.  Along the lines of~\cite{Appelquist_2009,Tekin:2010mm} we
interpolate our data for $\gbsf^2(\beta,L/a)$ and $\gbsf^2(\beta,2L/a)$. The
computation of the two-loop improved lattice step-scaling function
$\Sigsf^{(2)}(\usf,a/L)$~\cite{Bode:1999sm} thus requires $\gbsf^2(\beta,L/a)$
to be fixed at certain values of $\usf$ in line with~\eqref{eq:SFsteps}. The
continuum limit is taken individually at each value of $\usf_{m}$ with full
error propagation.  
The coarsest lattice only serves the purpose of estimating the size of cutoff
effects but is never taken into account in the analysis. Furthermore, data for
$2L/a=20$ is still missing and cannot be included at present. The plot provided
in Fig.~\ref{fig:CL} thus shows a (preliminary) continuum limit of
$\Sigsf^{(2)}(u,a/L)$ using a local fit ansatz with linear scaling in $(a/L)^2$
towards the continuum.  This gives the largest error on $\sigsf(u)$ which will
be improved after studying appropriate global fit ans\"atze in more detail.
With the data at hand we use the standard interpolation 
\begin{align}\label{eq:fit-sig}
     \sigma(u) &= u + s_0 u^2 + s_1 u^3 +s_2^{\rm fit} u^4 + s_3^{\rm fit} u^5 \;, &
           s_0 &= 2\ln(2) b_{0} \;,  &
           s_1 &= s_0^2 + 2 \ln(2)b_1 \;,
\end{align}
with parameters $s_2^{\rm fit},s_3^{\rm fit}$ and $s_0$, $s_1$ taken from
perturbation theory. In Fig.~\ref{fig:run_gSF} we plot the data points and
interpolating fit function of $\sigsf(u)$ together with their known
perturbative behaviour and remark that the important information lies in the
well-defined error of the former.  To look at our current results from a global
perspective, we compare it in Fig.~\ref{fig:strategy} (right panel) to previous
determinations of $\sigsf$ for $\nf=0,2,4$. For a better discrimination we plot
$[\sigsf(u)-u]/[2\ln(2)u^2]$ such that the axis intercept corresponds to $b_0$.
This figure provides a beautiful demonstration of our current understanding of
asymptotic freedom at the non-perturbative level.

\pagebreak
\section{Conclusions \& Outlook}
We have presented an extension to the ALPHA-strategy towards future high
precision computations of the Lambda parameter. It still keeps all systematic
errors under control by following a traditional finite-size scaling approach to
non-perturbatively connect low- and high-energy regimes. The gain in precision
is achieved by exploiting complementary properties of the SF and gradient flow
coupling together with a non-perturbative scheme switch at intermediate
energies.  While the SF part of the computations has been finished we are now
working on the computation of the gradient flow step-scaling function and the
non-perturbative scheme-switching step.

\section*{Acknowledgments}
\small
This work is supported in part by the Deutsche Forschungsgemeinschaft under
SFB/TR~9.  S. Sint acknowledges support by SFI under grant 11/RFP/PHY3218. 
M.D.B. is funded by the Irish Research Council, and is grateful for the
hospitality at DESY Zeuthen. We gratefully acknowledge the computer resources
provided by the John von Neumann Institute for Computing as well as at HLRN and
at DESY, Zeuthen.

\bibliography{mainbib}

\end{document}